\documentclass{iaus}
\usepackage{graphicx}

\title[Molecular processes AGB-PNe] 
{Molecular processes from the AGB to the PN stage}

\author[Garc\'{\i}a-Hern\'andez]   
{D. A. Garc\'{\i}a-Hern\'andez$^{1,2}$}

\affiliation{$^1$Instituto de Astrof\'{\i}sica de Canarias, V\'{\i}a L\'actea
s/n, E-38200 La Laguna, Spain \\ email: {\tt agarcia@iac.es} \\[\affilskip] $^2$
Departamento de Astrof\'{\i}sica, Universidad de La Laguna (ULL), E-38205 La
Laguna, Spain}

\pubyear{2012}
\volume{283}  
\pagerange{0-100}
\jname{Planetary Nebulae, an Eye to the Future}
\editors{A. Manchado, D. Schonberner, \& L. Stanghellini eds.}
\begin{document}

\maketitle

\begin{abstract}
Many complex organic molecules and inorganic solid-state compounds have been
observed in the circumstellar shell of stars (both C-rich and O-rich) in the
transition phase between Asymptotic Giant Branch (AGB) stars and Planetary
Nebulae (PNe). This short ($\sim$10$^{2}$-10$^{4}$ years) phase of stellar
evolution represents a wonderful laboratory for astrochemistry and provides
severe constraints on any model of gas-phase and solid-state chemistry. One of
the major challenges of present day astrophysics and astrochemistry is to
understand the formation pathways of these complex organic molecules and
inorganic solid-state compounds (e.g., polycyclic aromatic hydrocarbons,
fullerenes, and graphene in the case of a C-rich chemistry and oxides and
crystalline silicates in O-rich environments) in space. In this review, I
present an observational review of the molecular processes in the late stages of
stellar evolution with a special emphasis on the first detections of fullerenes
and graphene in PNe. 

\end{abstract}

\firstsection 
\section{Introduction}

At the end of the Asymptotic Giant Branch (AGB), low- and intermediate-mass
stars experience strong mass loss that efficiently enriches the interstellar
medium with gas and dust. The main processes of nucleosynthesis on the AGB are
the production of carbon and of heavy s-process elements. AGB stars experience a
different nucleosynthesis depending on the progenitor mass and metallicity
(e.g., Garc\'{\i}a-Hern\'andez et al. 2007a and references therein). In our own
Galaxy, low-mass (M$<$1.5 M$_{\odot}$) AGB stars remain O-rich and they probably
do not form a Planetary Nebula (PN), while intermediate-mass (1.5$<$M$<$4
M$_{\odot}$) turn C-rich and they produce s-process elements through the
$^{13}$C neutron source. Finally, high-mass (M$>$4 M$_{\odot}$) AGB stars remain
O-rich because of the Hot Bottom Burning (HBB) activation and they produce
s-elements via the $^{22}$Ne neutron source (Garc\'{\i}a-Hern\'andez et al.
2006, 2009). Note that this AGB evolutionary scenario is strongly modulated by
metallicity (e.g., 3$^{rd}$ dredge-up efficiency, HBB activation). The more
massive C-rich and O-rich sources are strongly obscured by the circumstellar
dust at the end of the AGB and they experience a phase of total obscuration in
their way to become PNe, being only accesible in the infrared and millimeter
wavelength ranges.

More than 60 molecules have been detected in the circumstellar shells around AGB
stars (e.g., Herbst \& van Dishoeck 2009). Many gas-phase molecules such as
inorganics (CO, SiO, SiS, NH$_{3}$, etc.), organics (C$_{2}$H$_{2}$, CH$_{4}$,
etc.), radicals (e.g., HCO$^{+}$), rings (e.g., C$_{3}$H$_{2}$), and chains
(e.g., HC$_{9}$N) have been detected around AGB stars. Gas-phase reactions
cannot  explain all of these molecules and solid-state chemistry has to be
considered (e.g., molecules can form on dust grains). Indeed, amorphous
silicates, weak crystalline silicates (olivine, pyroxene), refractory oxides
(corundum, spinel) have been detected around O-rich AGB stars. On the other
hand, SiC and amorphous carbon are frequently observed in
C-rich AGB stars, although other complex and disordered organic compounds (e.g.,
coal, kerogen, tholins) can provide the strong IR continuum and features
observed. However, young and evolved PNe show strong Aromatic Infrared Bands
(AIBs) - usually associated with Polycyclic Aromatic Hydrocarbons (PAHs) - and
crystalline silicates in the case of a C-rich and O-rich chemistry,
respectively. Mixed-chemistry (showing both C-rich and O-rich dust features) PNe
are also observed. The aromatic (C-rich) and crystalline (O-rich) compounds are
synthesized during the short ($\sim$10$^{2}$-10$^{4}$ years) transition phase
between AGB stars and PNe (e.g., Garc\'{\i}a-Lario \& Perea-Calder\'on 2003;
Kwok 2004). Thus, post-AGB stars and proto-PNe are wonderful laboratories for
astrochemistry, providing us with crucial information about the formation
pathways of complex organic molecules and inorganic solid-state compounds. Here
I present an observational review (both in C-rich and O-rich environments) of
the molecular processes (i.e., gas-phase molecules and the dust grains that
these molecules can form) from AGB stars to PNe as seen by the Infrared Space
Observatory (ISO) and the Spitzer Space Telescope, which give us strong
constraints to gas-phase and solid-state chemistry. Special attention is given
to the first detections of fullerenes and graphene in PNe.

\section{Molecular evolution from AGB to PNe as seen by ISO and Spitzer.}

{\underline{\it O-rich chemistry.}} ISO spectroscopy has revealed that weak
crystalline silicate (olivine, pyroxene) and water ice features are only seen in
high mass loss rate O-rich (C/O$<$1) AGB stars - the so-called OH/IR stars
(e.g., Sylvester et al. 1999; see Figure 1, left panel). Also, S-type AGB stars
(with C/O$\sim$1) show a set of O-rich infrared features that are different to
those seen in OH/IR stars (Hony et al. 2009). Another important result from ISO
was the detection of strong crystalline silicates in O-rich post-AGB stars and
PNe (e.g., Waters et al. 1996; Molster et al. 2002) as well as the
identification of double-dust chemistry objects (e.g., cool WCPNe, Waters et al.
1998). Water ice features (both amorphous and crystalline) were also detected in
some post-AGB stars (e.g., Manteiga et al. 2011 and references therein).

With the advent of the Spitzer Space telescope, it has been possible to study
O-rich AGB stars in other galaxies as well as to study the relation between the
dust properties and metallicity. It is found that O-rich AGBs are less obscured
(a lower dust production) in low metallicity environments and the amorphous
silicates are always seen in emission (e.g., Sargent et al. 2010). Another
important Spitzer result was the identification of the (heavily obscured)
high-mass precursors of PNe (the OHPNe). The latter sources showed unusual
crystalline silicate features, likely due to the different nucleosynthesis in
the previous AGB phase (Garc\'{\i}a-Hern\'andez et al. 2007b). In addition,
Spitzer has permitted to study the total obscuration phase that could not be
done by ISO (e.g., Garc\'{\i}a-Hern\'andez et al. 2007c). With regard to O-rich
PNe, Spitzer found that O-rich PNe are less common at low metallicity such as in
the Magellanic Clouds (MCs, Stanghellini et al. 2007). Also, many O-rich PNe
showing amorphous silicates emission have been detected by Spitzer (e.g.,
G\'orny et al. 2010; Stanghellini et al. 2011 and these proceedings).  

In short, the evolution of the O-rich dust features proceeds from amorphous (in
the AGB phase) to crystalline silicates (in the PNe stage) (Garc\'{\i}a-Lario \&
Perea-Calder\'on 2003). Two models for the crystallization of the silicates have
been proposed: i) high-temperature crystallization at the end of the AGB as a
consequence of the strong mass loss (Waters et al. 1996) or ii) low-temperature
crystallization in long-lived circumbinary disks (Molster et al. 1999).

\begin{figure}[b]

\begin{center}
 \includegraphics[width=2.6in]{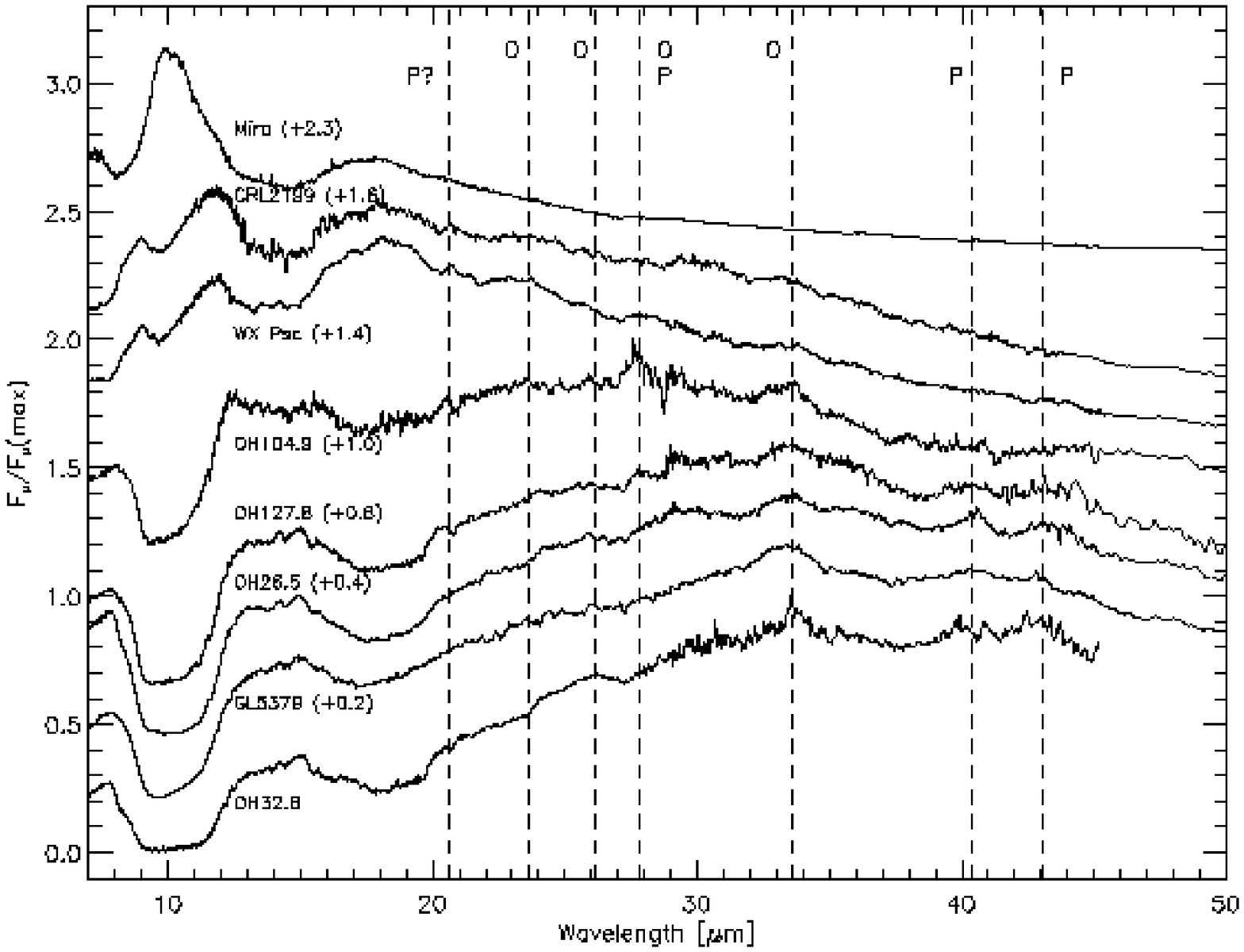}%
 \includegraphics[width=2.8in]{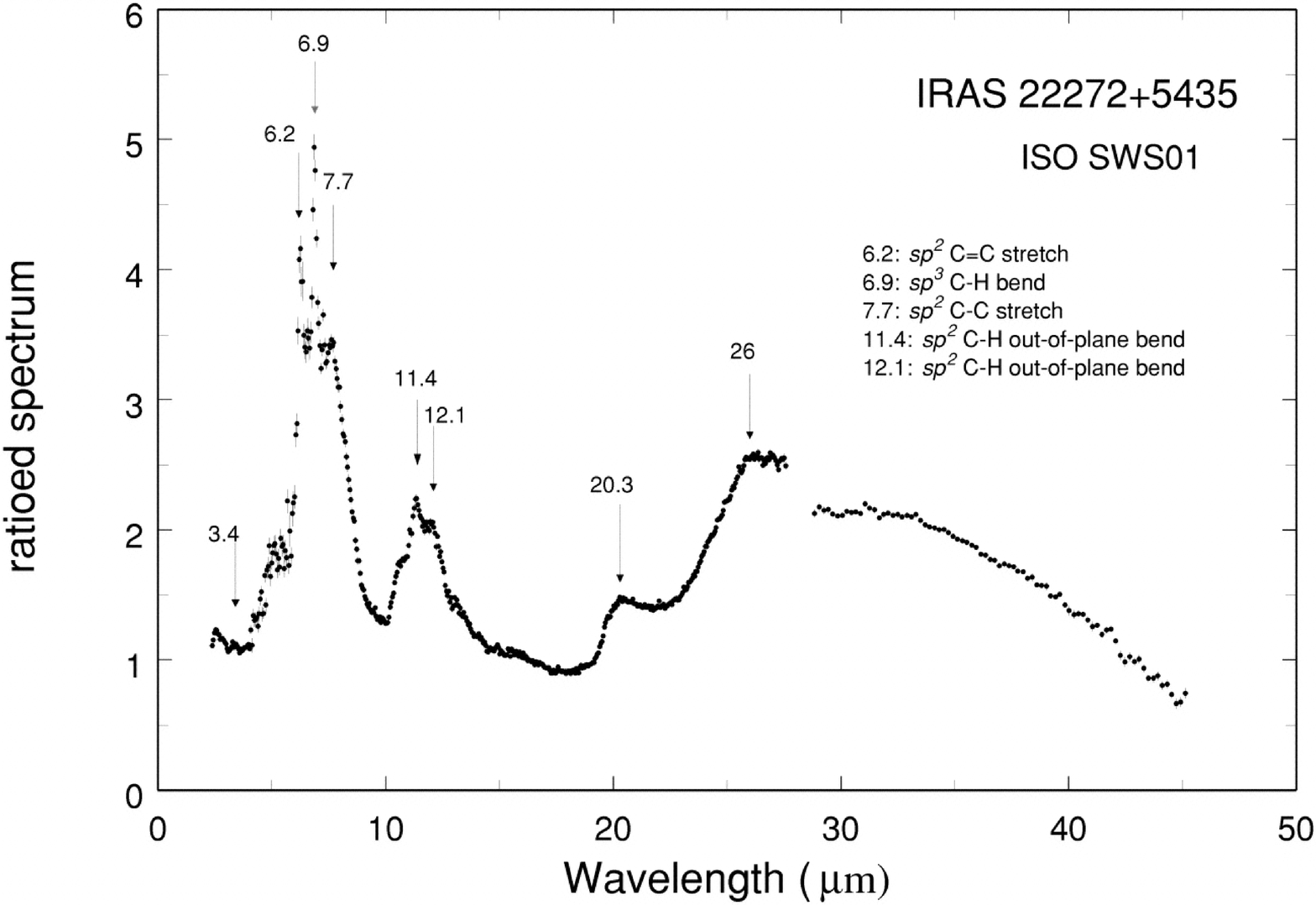} 
 \caption{{\it Left panel}: Weak crystalline silicate features from pyroxene (P)
and olivine (O) found in OH/IR stars by ISO (Sylvester et al. 1999). {\it Right
panel}: ISO spectrum of the proto-PN IRAS 22272$+$5435 showing the presence of
aliphatic discrete features (e.g., at 3.4 and 6.9 $\mu$m) and the 8 and 12
$\mu$m aliphatic plateaus together with AIBs (e.g., at 6.2 and 7.7 $\mu$m). The
still unidentified 21, 26, and 30 $\mu$m features are also seen (Kwok et al.
2001).}
   \label{fig1}
\end{center}
\end{figure}

{\underline{\it C-rich chemistry.}} Gas-phase organic molecules (e.g.,
C$_{2}$H$_{2}$, HCN, etc.) around the prototype C-rich AGB star IRC $+$10216
were first detected by ISO (Cernicharo et al. 1999). These organic molecules are
usually observed together with a strong dust continuum emission and a broad dust
feature centered at $\sim$11.5 $\mu$m (e.g., Yang et al. 2004). The strong dust
continuum emission is usually attributed to amorphous carbon while the 11.5
$\mu$m feature is believed to be produced by SiC (e.g., Speck et al. 2009). An
unidentified broad feature at 25-35 $\mu$m (the so-called 30 $\mu$m feature;
e.g., Volk et al. 2000) is already seen during the AGB phase. The 11.5 and 30
$\mu$m features are also observed in the post-AGB and PNe phases (e.g., Hony et
al. 2002; Morisset et al. these proceedings) but the still unidentified 21
$\mu$m feature is genuine of the post-AGB stage (see Sect. 3).

The ISO detection of other C-bearing species such as polyynes, benzene, etc. in
post-AGB stars may be explained by the polymerization of C$_{2}$H$_{2}$, HCN,
and carbon chains (Cernicharo 2004). Indeed, these small hydrocarbon molecules
like C$_{2}$H$_{2}$, C$_{4}$H$_{2}$, C$_{6}$H$_{6}$ have been suggested to be
the building blocks of more complex molecules such as PAHs and that are known to
show strong aromatic infrared bands (AIBs at 3.3, 6.2, 7.7, 8.6, and 11.3
$\mu$m; e.g., Allamandola et al. 1989) coincident with the unidentified infrared
(UIR) emission observed in stars evolving from AGB stars to PNe. However,
because of the low temperatures of the central stars, it is difficult to believe
that the UIRs observed in proto-PNe are due to free-flying gas-phase PAHs (e.g.,
Kwok et al. 2001; Duley \& Williams 2011). Interestingly, proto-PNe show
aliphatic emission represented by the 8 and 12 $\mu$m emission plateaus and the
3.4, 6.9, and 7.3 $\mu$m features together with the classical AIBs (see Figure
1, right panel). This is a strong indication of the coexistence of aliphatic and
aromatic structures in the circumstellar shells of these evolved stars. Thus, a
better alternative to the PAH hypothesis is a solid material with a mix of
aliphatic and aromatic structures such as hydrogenated amorphous carbon (HAC;
e.g., Duley \& Williams 2011), coal (e.g., Guillois et al. 1996), etc.
Observationally, it seems clear that the aliphatic component decreases with the
evolutionary stage, with the AIBs being stronger in the PN phase (Kwok et al.
2001; Garc\'{\i}a-Lario \& Perea-Calder\'on 2003). This change from aliphatic to
aromatic structures was attributed to the photochemical processing by the UV
photons from the central star (Kwok et al. 2001). 

Our understanding of the C-rich chemistry in evolved stars has significantly
improved with the more recent Spitzer data, permitting us to study extragalactic
sources in the transition phase from AGB stars to PNe for the first time.
Infrared features from gas phase molecules (C$_{2}$H$_{2}$, HCN, C$_{3}$) are
more common and strong in AGB stars of low metallicity environments (e.g.,
Lagadec et al. 2007). Examples of truly extragalactic C-rich post-AGB stars were
first detected by Bernard-Salas et al. (2006). More recently, Volk et al. (2011)
have reported the Spitzer spectra of a significant sample of post-AGB stars in
the MCs, finding that the unidentified 21 $\mu$m feature is more
common at low metallicity. The systematic study of heavily obscured post-AGB
stars of our Galaxy show that aliphatic broad emissions (at $\sim$10-15, 15-20,
25-35 $\mu$m) as well as molecular absorptions from small hydrocarbons are
typical during the total obscuration phase of the more massive C-rich post-AGB
stars (Garc\'{\i}a-Hern\'andez et al. 2007c). As expected from the
nucleosynthesis in the previous AGB phase, C-rich PNe were found to be more
frequent at low metallicity, showing less processed dust grains (i.e., aliphatic
dust dominates and AIBs are rare; Stanghellini et al. 2007).

In summary, the C-rich dust features seem to change from aliphatic (in the AGB
phase) to mostly aromatic (in the PN stage) although a mix of aliphatic and
aromatic features are observed in the post-AGB stage and still in some PNe.
Several scenarios have been proposed to explain the aromatization from the AGB
phase to the PNe stage: i) dust processing (change from aliphatic to aromatic
structures) by the UV photons from the central star (e.g., Kwok et al. 2001;
Garc\'{\i}a-Lario \& Perea-Calder\'on 2003; Kwok 2004); ii) acetylene
(C$_{2}$H$_{2}$) and its radical derivatives are the precursors of PAHs (e.g.,
Cernicharo 2004). The very recent detections of fullerenes and graphene in
post-AGB stars and PNe have provided new valuable information about the dust
processing in the circumstellar shells of evolved stars (see Sect. 4). 

{\underline{\it Mixed-chemistry sources.}} The mixed chemistry phenomenon (the
simultaneous presence of both C-rich and O-rich chemistry and dust discovered by
ISO) was found to be a common characteristic to cool Wolf-Rayet WCPNe, pointing
to the presence of a long-lived circumbinary disk (Waters et al. 1998). More
recent Spitzer observations show that the mixed-chemistry phenomenon is more
common in the Galactic Bulge, being not restricted to cool WCPNe
(Perea-Calder\'on et al. 2009). The binary hypothesis cannot explain the
dual-dust chemistry phenomenon in the Galactic Bulge and a very late thermal
pulse (Perea-Calder\'on et al. 2009) or hydrocarbon chemistry in an
UV-irradiated, dense torus (Guzm\'an-Ram\'{\i}rez et al. these proceedings) have
been invoked to explain the high detection rate of mixed-chemistry sources in
the Bulge. In addition, mixed-chemistry has now been detected in some post-AGB
stars (Cerrigone et al. 2009). Interestingly, the detection rate of
mixed-chemistry PNe is strongly linked to the metallicity (Stanghellini et al.
2011 and these proceedings).

\section{The unidentified 21, 26, and 30 $\mu$m features}

Apart from the aromatic and aliphatic features mentioned above, there is an
interesting set of still UIR features located at 21, 26, and 30 $\mu$m and that
are usually observed in stars from the AGB to the PN phase. 

{\underline{\it The 21 $\mu$m feature.}} This feature is only observed in
post-AGB stars and their characteristics indicate a solid-state carrier with a
fragile nature. Many different carriers of the 21 $\mu$m feature have been
proposed in the literature but it seems clear that the carrier should be a
carbonaceous compound. Hydrogenated fullerenes, nanodiamonds, HAC, TiC
nanoclusters, etc. are some examples among the proposed carbonaceous species.
Amides (urea or thiourea) as the carrier of the 21 $\mu$m emission is the more
recent and exotic proposal (Papoular 2011). 

{\underline{\it The 26 and 30 $\mu$m features.}} The 30 $\mu$m feature,
sometimes with substructure at 26 $\mu$m, is seen from the AGB to the PN stage
with an important fraction of the total energy output. This indicates that the
carrier should be very abundant in the circumstellar shell. The carrier of the
30 $\mu$m feature is usually assumed to be MgS (Hony et al. 2002) but Zhang et
al. (2009) argue that MgS is very unlikely the carrier of the 30 $\mu$m emission
seen in C-rich evolved stars. Grishko et al. (2001) also show that HACs can
explain the 30 $\mu$m feature. The HAC identification can also explain the 21
and 26 $\mu$m features. More recently, Papoular (2011) shows that aliphatic chains (CH$_{2}$ groups, O
bridges and OH groups) can also explain the 30 $\mu$m emission in sources
evolving from AGB to PNe. Finally, it should be noted that other weaker UIR
features at 15.8, 16.4, and 17 $\mu$m are also seen in some proto-PNe (see e.g.,
Hrivnak et al. 2009).

\section{Fullerenes and graphene in circumstellar envelopes.}

Fullerenes such as C$_{60}$ and C$_{70}$ are highly resistant and stable
tridimensional molecules formed exclusively by carbon atoms. Fullerenes and
fullerene-related molecules have attracted much attention since their discovery
at laboratory (Kroto et al. 1985) because they may explain certain unidentified
astronomical features such as the so-called diffuse interstellar bands (DIBs)
(see e.g., Luna et al. 2008 for a review on interstellar/circumstellar DIBs).
The remarkable stability of fullerenes against intense radiation, ionization,
etc. reinforced the idea that fullerenes should be present in the interstellar
medium with important implications for interstellar/circumstellar chemistry.
Indeed, fullerenes were found on Earth an on meteorites and several unsuccessful
ISO attempts to look for the mid-IR signatures of fullerenes in circumstellar
shells have been previously made, including R Coronae Borealis (RCB) stars
(e.g., Lambert et al. 2001) and post-AGB stars (e.g., Kwok et al. 1999).

{\underline{\it Discovery of fullerenes in space.}} The RCB stars have been
considered to be the ideal astrophysical environments for the formation of
fullerenes (Goeres \& Sedlmayr 1992). This is because the H-deficiency together
with the He and C-rich characters of RCBs resemble the experimental conditions
on Earth, facilitating the formation of fullerenes (e.g., Kroto et al. 85). In
late 2009 Garc\'{\i}a-Hern\'andez, Rao \& Lambert looked for C$_{60}$ in a
complete sample of 31 RCBs using Spitzer. They got the unexpected result that
C$_{60}$ was only detected in the two RCBs (DY Cen and V854 Cen) with some H in
their circumstellar shells. Interestingly, their detection of C$_{60}$ around
RCBs occurred in conjunction with the presence of PAHs. In addition, they found
that the V854 Cen's IR spectrum evolved from HACs (ISO 1996's spectrum) to PAHs
and C$_{60}$ (Spitzer's 2007) (see Figure 2, left panel). These unique IR
spectral variations prompted them to claim that the PAHs and fullerenes evolved
from HAC grains. Because of the unexpected result in RCBs, they looked for
fullerenes in $\sim$240 PNe pertaining to very different environments and
observed by Spitzer. The mid-IR signatures of the C$_{60}$ fullerenes were
clearly found in five PNe with normal H abundances (including the PN Tc 1). The
common presence of PAH features in fullerene-containing PNe confirmed them the
unexpected results obtained in RCBs. Meanwhile, Cami et al. (2010) reported in
{\it Science} the great discovery of the IR detection of C$_{60}$ and C$_{70}$
fullerenes in Tc 1 as due to the H-poor conditions in the inner
core\footnote{Garc\'{\i}a-Hern\'andez, Rao \& Lambert submitted their unexpected
RCB results to {\it Science} on April 19th 2010 but their work was rejected;
e.g., one reviewer was opposed to their work because it was {\it ``in the wrong
direction with the least H-deficient stars showing the C$_{60}$ features"}.}.
However, neither the central star, nor the inner core and the nebula are
H-deficient and current understanding of stellar astrophysics does not allow for
Tc 1 being a H-poor PN (see Garc\'{\i}a-Hern\'andez et al. 2010;
Garc\'{\i}a-Hern\'andez, Rao \& Lambert 2011). Thus, the detection of fullerenes
in PNe and RCB stars have challenged the paradigm regarding fullerene formation
in space, showing that, contrary to general expectation, fullerenes are
efficiently formed in H-rich circumstellar environments. In addition, C$_{60}$
has been recently detected in the proto-PN IRAS 01005$+$7910 (Zhang \& Kwok
2011) and two binary post-AGB stars (Gielen et al. these proceedings),
indicating that fullerene formation can occur just after the AGB phase.

\begin{figure}[b]
\begin{center}
\includegraphics[width=2.6in]{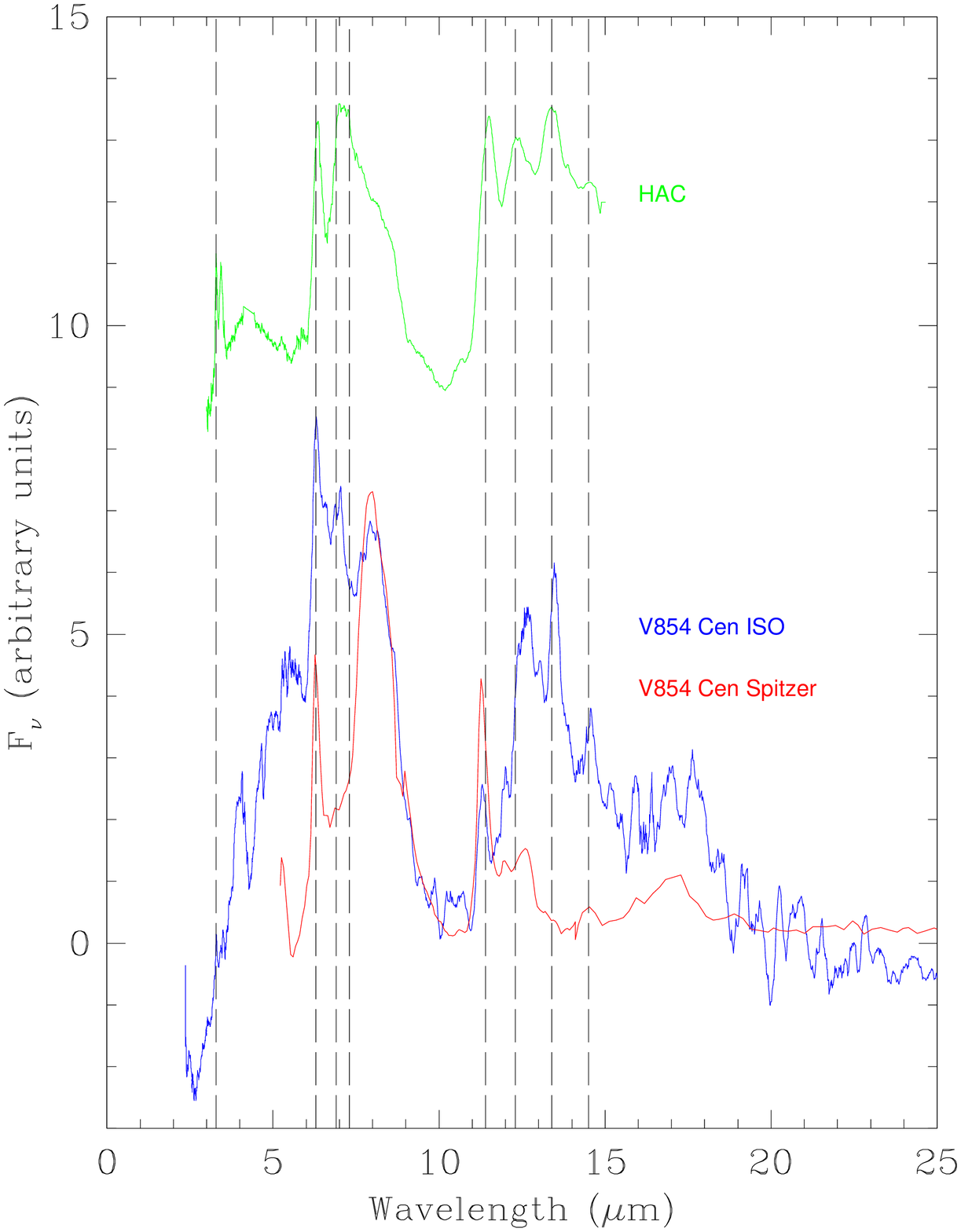}%
\includegraphics[width=2.6in]{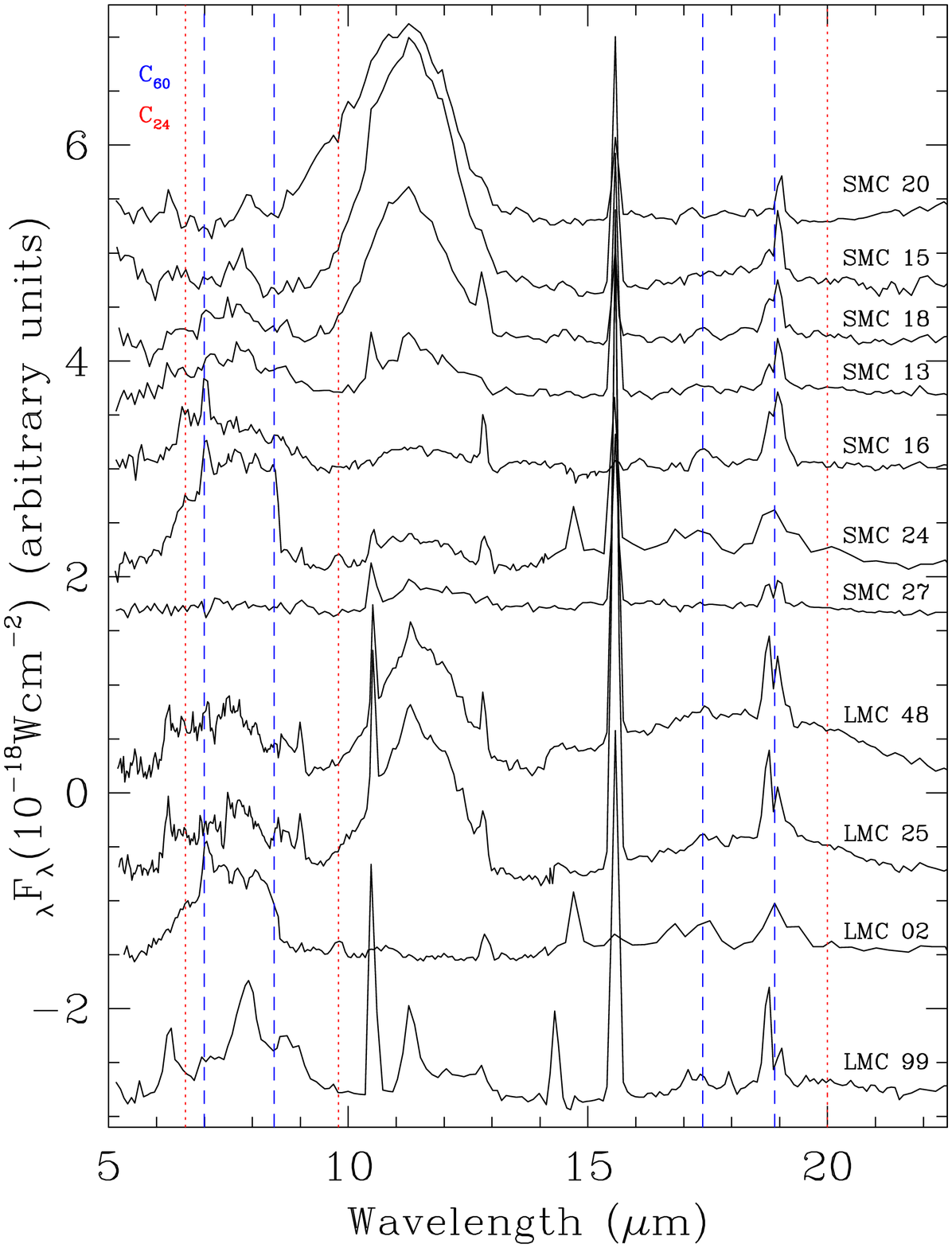} 
\caption{{\it Left panel}: Residual ISO and Spitzer/IRS  spectra for the RCB
star V854 Cen. The laboratory emission spectrum of HAC at 773 K is shown for
comparison (Garc\'{\i}a-Hern\'andez, Rao \& Lambert 2011). {\it Right panel}:
Residual Spitzer spectra of fullerene-MCPNe. The band positions of C$_{60}$
(dashed) and planar C$_{24}$ (dotted) are marked (Garc\'{\i}a-Hern\'andez et al.
2011).}
\label{fig1}
\end{center}
\end{figure}

{\underline{\it Formation of fullerenes in H-rich ejecta.}} The simultaneous
detection of PAH-like features and fullerenes toward C-rich and H-containing PNe
indicates that modifications are needed to our current understanding of the
chemistry of large organic molecules as well as the chemical processing in space
(Garc\'{\i}a-Hern\'andez et al. 2010). The suggestion was made that both
fullerenes and PAHs can be formed by the photochemical processing (e.g., as a
consequence of UV irradiation or shocks) of HAC, which should be a major
constituent in the circumstellar envelope of C-rich evolved stars. This idea is
supported by the laboratory experiments on the decomposition of HAC, which show
that the products of destruction of HAC grains are PAHs and fullerenes in the
form of C$_{50}$, C$_{60}$, and C$_{70}$ molecules (Scott et al. 1997). More
recently, Garc\'{\i}a-Hern\'andez et al. (2011) have presented new Spitzer
detections of C$_{60}$ and C$_{70}$ fullerenes in PNe of the MCs
(MCPNe), which have permitted an accurate determination of the C$_{60}$ and
C$_{70}$ abundances (C$_{60}$/C$\sim$0.07\% and C$_{70}$/C$\sim$0.03\%) in space
for the first time. The quantitative determination of fullerenes in space was
possible thanks to the very recent laboratory studies of the C$_{60}$ and
C$_{70}$  fullerenes (Iglesias-Groth et al. 2011; Cataldo et al. these
proceedings). In addition, the new MCPNe studies show that neutral fullerenes
are likely in solid-state and collisionaly excited (i.e., they are not excited
by the UV photons from the central stars) and that they probably evolved from
the shock-induced decomposition of small solid particles similar to that of
HAC dust (Garc\'{\i}a-Hern\'andez et al. 2011).

{\underline{\it First detection of graphene in space.}} Interestingly, some of
the fullerene-containing MCPNe display very unusual emission features at
$\sim$6.6, 9.8, and 20 $\mu$m (see Figure 2, right panel) coincident with the
theoretical transitions of planar C$_{24}$ (Garc\'{\i}a-Hern\'andez et al.
2011). Planar C$_{24}$ is a very stable molecule (more than the C$_{24}$
fullerene) and can be viewed as a small fragment of a graphene sheet. The
detection of these very unusual features in fullerene sources represents the
first possible detection of graphene in space. However, a definitive
confirmation has to wait for laboratory infrared spectroscopy of C$_{24}$, which
is extremely difficult because of the high reactivity of C$_{24}$. The possible
detection of graphene precursors (C$_{24}$) opens the possibility of detecting
other complex forms of carbon (e.g., carbon nanotubes, carbon onions, etc.) in
space.

{\underline{\it UIR emissions in fullerene sources.}} Remarkably, fullerenes and
graphene have been detected in PNe whose IR spectra are  dominated by aliphatic
C-rich dust, represented by broad emissions such as those at 6-9, 10-15, 15-20,
and 25-35 $\mu$m. The 6-9 $\mu$m feature may be attributed to HACs or large PAH
clusters. On the other hand, the broad 10-15 $\mu$m (the 11.5 $\mu$m feature)
and the 25-35 $\mu$m emission (the 30 $\mu$m feature) are usually attributed to
SiC (e.g., Speck et al. 2009) and MgS (e.g., Hony et al. 2002), respectively.
However, the observed variability of these broad features is quite consistent
with the variable properties of HACs (a material with mixed aliphatic and
aromatic structures), which are able to provide a wide range of different
spectra (e.g., the relative strength and position of the IR features) depending
on their physical and chemical properties (e.g., size, shape, hydrogenation;
e.g., Grishko et al. 2001). Indeed, these aliphatic emissions in low-metallicity
extragalactic post-AGB stars and PNe are stronger and much more common than in
the higher metallicity Galactic counterparts (Volk et al. 2011; Stanghellini et
al. 2007, 2011). Because of the lower metal content (Si, Mg) of MC post-AGBs and
PNe, the opposite is expected. Thus, it is very unlikely that the carriers of
the broad 11.5 and 30 $\mu$m emissions - usually seen in fullerene-containing
PNe - are related with SiC and MgS, respectively, as it was suggested in the
past. The carriers of the broad 11.5 and 30 $\mu$m features (also the bump at
15-20 $\mu$m and possibly the so-called 21 and 26 $\mu$m features) are more
likely related with other decomposition products of HACs or a similar material
(e.g., fullerene and graphene precursors or intermediate products not yet
identified) but a definitive answer requires further laboratory efforts. 

\section{Concluding remarks}

In summary, the most likely explanation for the formation of fullerenes and
graphene in H-rich environments is that these molecular species may be formed
from the destruction of a carbonaceous compound with a mixture of aromatic and
aliphatic structures - e.g., HAC - which should be widespread in space. In this
context, the coexistence of a large variety of molecular species in H-rich
circumstellar environments is supportive of a model in which non-equilibrium IR
emission occurs from small solid particles containing aromatic, aliphatic,
fullerene, and graphene structures similar to that of HAC dust. Indeed, Duley \&
Williams (2011) have recently suggested a model for the heating of HAC dust via
the release of chemical energy that gives a natural explanation for the
astronomical aromatic emission at 3.3 (and other UIR wavelengths) usually
attributed to PAHs. Finally, it should be noted here that instead of HACs, other
materials with a complex mix of aromatic and aliphatic structures (e.g., coal,
kerogen, petroleum fractions, soot, quenched carbonaceous composites, carbon
nanoparticles, etc.) can be present in astrophysical environments and they
should be widely explored at laboratory. For example, coal and petroleum
fractions may explain the great diversity  of spectral features (aromatic and
aliphatic) seen in proto-PNe (Guillois et al. 1996; Cataldo et al. 2002).

\end{document}